# Switching of controlling mechanisms during the rapid solidification of a melt pool in additive manufacturing

Yijia Gu[1,*] and Lianyi Chen[2,3]

1. Department of Materials Science and Engineering, Missouri University of Science and Technology, Rolla, MO, 65409, USA
2. Department of Mechanical Engineering, University of Wisconsin-Madison, Madison, WI, 53706, USA
3. Department of Materials Science and Engineering, University of Wisconsin-Madison, Madison, WI, 53706, USA

**ABSTRACT**

Fusion-based metal additive manufacturing (AM) is a disruptive technology that can be employed to fabricate metallic component of near-net-shape with an unprecedented combination of superior properties. However, the interrelationship between AM processing and the resulting microstructures is still not well understood. This poses a grand challenge in controlling the development of microstructures during AM to achieve desired properties. Here we study the microstructure development of a single melt pool, the building block of AM-fabricated metallic component, using a phase-field model specifically developed for the rapid solidification of AM. It is found that during the rapid solidification of the melt pool, the solid-liquid interface is initially controlled by solute diffusion followed by a thermal diffusion-controlled stage with an undercooling larger than the freezing range. This switching of controlling mechanisms leads to the sudden changes in interfacial velocity, solute concentration, and temperature, which perfectly explains the formation of various heterogeneous microstructures observed in AM. By manipulating the processing conditions, the switching of controlling mechanisms can be controlled to form refined microstructures or layered structures for improved mechanical properties and resistance to cracking.

## 1. INTRODUCTION

Fusion-based metal additive manufacturing (AM) is capable of fabricate metallic component of near net shape directly from a digital file[1]. Since the fusion-based metal AM exploits the far-from-equilibrium process of rapid solidification, it enables the development of novel microstructures with extended solubility, reduced partitioning of solute, novel metastable phases, and refined structures[2], and hence opens up a new horizon of microstructure synthesis to achieve unique properties or a unique combination of them. In addition, AM promises the fabrication of materials with site-specific properties[3] or functionally graded materials[4] by tuning the microstructures from point to point. Thus, AM shows the potential of producing novel metal components fueled by locally-tailored superior properties, and may lead to the paradigm shift from "design for manufacturing" to "manufacturing for design"[5].

Nevertheless, the interrelationship between AM processing and microstructure development is not well understood. The AM processing parameters for a given alloy are largely determined by trial and error. In addition, almost all the alloys currently being used by AM are designed for conventional processes, which cannot produce desired microstructures and often results in many incompatibility issues such as cracking and inferior properties[6]. To fundamentally resolve those issues and more importantly to unleash the full potential of AM, the relationship between AM processing and microstructure development needs to be established.

The structure of AM-fabricated metallic component is made up of repeating melt pools with designed orientation and overlapping. Understanding the mechanism of microstructure development inside the melt pool during its solidification process is the key. However, the microstructures developed in the melt pool are not uniform. Many even show complex heterogeneous structures, which include the formation of the coarse band at the bottom of the melt pool in AlSi10Mg[7,8], the development of bimodal grain structures in Al alloys with excess Sc and/or Zr[9–15], the formation of eutectic structures in hypoeutectic Al-Si alloys[16], and many others[17].

The microstructure of each melt pool develops via the so-called rapid solidification, a far-from-equilibrium process, due to the high cooling rate. The resulting microstructures are determined by the complex interplay between interfacial velocity, temperature profile, solute petitioning, and interface morphology during the solidification process. The development of microstructure in a rapid solidification process can be understood structurally and compositionally[18]. Structurally, the morphology of the solid-liquid interface which includes planar, cellular, and dendritic, depends on the thermal gradient ($G = dT/dx$) at the solidification front and the growth rate or interfacial velocity $v_\mathrm{i}$. The microstructural features, such as the cell size and dendrite arm spacing, are strongly dependent on the cooling rate ($\dot{T} = dT/dt$). Compositionally, the solute distribution across the solid-liquid interface, or partitioning, becomes strongly dependent on interfacial velocity due to the breakdown of local equilibrium at the interface. Higher interfacial velocity leads to larger partition coefficient and stronger solute trapping effect. The evolution of temperature profile during rapid solidification is complicated by the release and dissipation of latent heat. The release rate of latent heat depends on $v_\mathrm{i}$. The dissipation rate of latent heat depends on thermal diffusion, which is typically assumed to be infinite fast for conventional casting but should be considered as finite for rapid solidification[19]. Therefore, to fully understand the microstructure development in AM, it is required to capture the evolution of interfacial velocity, temperature profile and solute distribution across the whole melt pool.

Our current understanding is still restricted due to limited access to the evolution of all those variables of the entire melt pool during the whole solidification process. Experimentally, the velocity of the solid-liquid interface during AM process can be measured using in situ X-ray imaging[20] or in situ transmission electron microscopy[21–24]. However, the temperature evolution of the interface is inaccessible due to the spatial and temporal resolution limit of current characterization techniques. (See Section 5 of Supplementary Materials for example.) On the modeling side, the solute diffusion and thermal diffusion are decoupled in most computational models. Many finite element method (FEM) packages can predict the temperature evolution of the melt pool very well, but they usually ignore the solute partitioning and solute diffusion, which are critical for the initial transient of the solidification process. Therefore, they are suitable for predicting the formation of liquid melt pool but not the solidification of it. Several phase-field models have been developed to study the rapid solidification[25–37]. Nevertheless, most of the current phase-field studies of AM processes decoupled the thermal diffusion from solute partitioning by imposing a fixed temperature profile either with a fixed thermal gradient[38] or taken from FEM simulations[39,40]. Those treatments may be adequate to simulate evolution of grain structures or the phenomena of a larger scale. But they are inadequate to simulate the sub-grain scale microstructure development due to insufficient description of the interplay between interfacial velocity, thermal profile, and solute partitioning. For instance, the local recalescence and the resulting variations of velocity and solute partitioning cannot be captured.

In this work, we employed a phase-field model[19] incorporated with coupled solutal-thermal diffusion and solute trapping effect to explore the microstructure development in the rapid solidification of a melt pool. We assume the melt pool has a perfect hemisphere geometry such that the simulation can be simplified to one-dimensional (1D) using a spherical coordinate system. This setup enables us to simulate the complete solidification process of the entire melt pool, while maintaining adequate resolution to capture the dynamics of the solid-liquid interface. The phase-field simulations, although much simplified, successfully captured the interplay between interfacial velocity, solute partitioning, and thermal diffusion. More importantly, the simulation results lead to the discovery of the switching of the controlling mechanisms of the microstructure development during the rapid solidification of the melt pool, which has not been reported before. The main focus of this communication is to explore the nature of the switching of the controlling mechanisms and to understand how it leads to the formation of various unexpected heterogeneous microstructures observed in AM experiments.

**2. RESULTS AND DISCUSSION**

**2.1 The Rapid Solidification of a Melt Pool of Hemisphere Shape**

To investigate the development of microstructures during AM, we simulated the solidification of a melt pool of Al-10Si as a simplified version of AlSi10Mg using the developed 1D phase-field model[19]. AlSi10Mg is one of the few AM-compatible light-weight alloys that has been widely studied[6]. For simplicity, the formation of secondary phases is not considered, since the skeleton of the formed microstructure, which is the focus of this work, is determined by the primary $\alpha$-Al phase. Due to the fast thermal diffusion, the temperature profile, which is complex across the whole melt pool, evolves quickly[19]. Therefore, the simulation needs to cover the whole melt pool to correctly capture the temperature evolution. We assume the melt pool has the geometry of a hemisphere such that the whole melt pool can be simulated in 1D using spherical coordinates. As illustrated in the inset of Fig. 1f, the thermal boundary condition of the right end, which is the top center of the melt pool, is no flux. On the other end of the simulation domain, which is the boundary of the melt pool, an interfacial thermal conduction boundary condition is enforced, i.e., $-K\frac{\partial T}{\partial x} = -h(T - T_\infty)$, where $K = \alpha c_p$ is the thermal conductivity, $h = 1.5 \times 10^6 \; Wm^{-2}K^{-1}$ is the heat transfer coefficient[41], and $T_\infty = 300 \; K$ is the temperature of the substrate. No flux boundary condition is applied to both $\theta$ (normalized temperature) and $U$ (normalized composition). The simulation starts with a uniform temperature of 1200 K and uniform composition $c_\infty = 10 \; wt\%$ Si. A tiny portion at the melt pool boundary is set to be solid ($\phi = 1$), and the rest of the simulation domain is set as liquid ($\phi = -1$). The simulation completes when 99% simulation domain is solidified. Although this simulation simplified the conditions of the solidification of a melt pool, it captured the evolution of temperature, solute concentration, and interfacial velocity during the solidification of the whole melt pool, which are essential for understanding the formation mechanism of the microstructures in AM. In the following, we discuss the unique findings revealed by this simulation.

Firstly, the solidification of the simulated melt pool consists of a low velocity stage and a high velocity stage (middle panel of Fig. 1f). In the beginning of the first stage, the temperature decreases dramatically, while the velocity of the interface remains low (~0.01m/s). When the interfacial temperature drops below a critical temperature, the interfacial velocity increases dramatically (see the sharp increase of the velocity in the middle panel of Fig. 1f). The sudden increase of the velocity leads to the rapid release of latent heat and consequently the increase of interfacial temperature, i.e., the recalescence. Then, the solidification enters the second stage with much higher velocity and relative constant temperature. The predicted interface velocity of this stage (0.1~0.3 m/s) is similar to experimental measurement from a selected laser melting (SLM) study on AlSi10Mg[42]. The rapid increase in velocity is expected to cause a sudden refining of the microstructures and hence form a sharp boundary separating coarse and fine structures. The width of the coarse structures developed in the initial stage depends on the cooling rate enforced at the initial boundary of the melt pool. For this simulation, the width is ~20 μm. Experimentally, it was observed that in AlSi10Mg there is a coarse band of 5 to 20 μm wide and the cell size inside the coarse band is about 4 times that of the cells in the rest of the melt pool[7,8]. Fig. 1b shows similar coarse band in Al-11.28Si alloy processed by directed energy deposition[16]. Our predictions agree with the structural features observed experimentally at least semi-quantitatively.

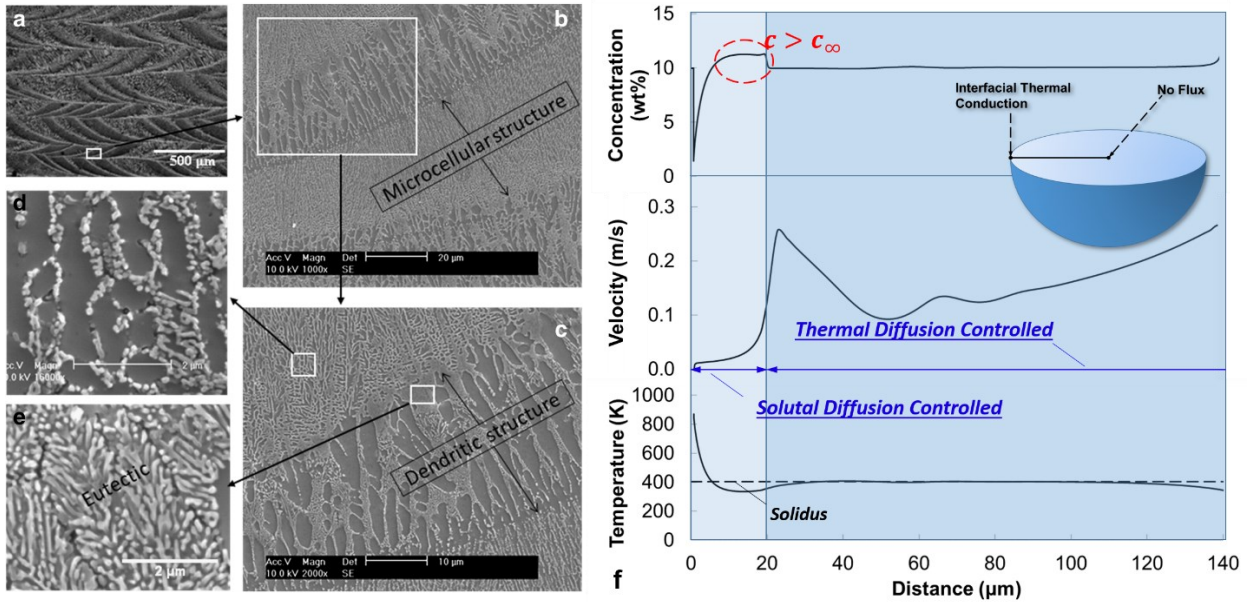

Figure 1. Microstructure development of melt pools for hypoeutectic Al-Si alloy with composition close to the eutectic point. (a-e) SEM micrographs showing the columnar dendrites and microcellular morphology at different points of the sample in laser deposited Al-11.28%Si [16]. (f) The phase-field simulation of Al-10%Si. The calculated instantaneous concentration (wt%) of the solid side of the interface, interfacial temperature, and interfacial velocity are plotted as functions of distance to the melt pool boundary. The inset is the setup of the phase-field simulation of a hemisphere melt pool in 1D spherical coordinates with interfacial thermal conduction on the left boundary and non-flux boundary condition on the right.

Secondly, the phase-field simulation predicted a composition variation in the resulting solid. Due to the decrease of temperature, the interfacial velocity $v_i$ kept increasing during the initial stage, which lead to the decrease of the characteristic length of solute diffusion ($l_0 = D_l/v$). Consequently, the solute in the boundary layer kept increasing and resulted in a solute-rich band[43]. As indicated in the top panel of Fig 1f, the predicted solute concentration becomes higher than the alloy composition $c_\infty$ as the solidification approaches the end of the first stage (or getting close to the boundary of the coarse band). Therefore, for hypoeutectic alloys with compositions close to the eutectic point, eutectic structures may form near the boundary between the coarse and fine structures. This is also confirmed experimentally in a laser deposited Al-11.28%Si alloy as shown in Fig. 1c and e. [16] (The eutectic point of Al-Si system is 12.6wt%Si.)

It should be noted that the anisotropy of the interface cannot be considered since this is a 1D model. The effect of interfacial anisotropy, or the curvature effect, can be quantified as the curvature undercooling. In Supplementary Materials, we show that the curvature undercooling has moderate contribution to the overall undercooling when the velocity of the interface is low or when the overall undercooling is low. But as the overall undercooling approaches the freezing range and beyond, its contribution diminishes and becomes negligible. Therefore, even though the curvature effect is not considered, our simulations agree quite well with experimental observations.

In summary, the results from 1D phase-field modeling agree well with experimental observations structurally and compositionally, although the simulation conditions are much simplified. As shown in Fig. 1f, the calculated temperature profile is rather complicated, which cannot be achieved without considering the dissipation of latent heat. In addition, it can also be inferred from Fig. 1f that the dependence of interfacial velocity on undercooling is complex. Therefore, the coupled thermal-solutal diffusion needs to be incorporated to make quantitative predictions when modeling the rapid solidification of the melt pool.

## 2.2 The Controlling Mechanism

As shown in the lower panel of Fig. 1f, the interfacial temperature in the second stage is barely below the solidus, which indicates the system is in the single-phase region on a binary phase diagram. Correspondingly, the interfacial velocity $v_i$ starts to increase rapidly when the interfacial temperature is below solidus as shown in the middle panel of Fig. 1f. Therefore, we hypothesize that *the sudden velocity increase is caused by the crossing of the solidus, i.e., moving from the solid-liquid two-phase region to the single solid phase region on the equilibrium phase diagram*.

Since the solidification of the melt pool starts from the bottom solid of the previous layers with a velocity of zero, the initial solidification can be regarded as at equilibrium, which is confirmed by the calculated initial composition and interfacial velocity by the phase-field simulation (Fig. 1f). At this point, both solid and liquid phase exist, and the system lies in the two-phase region. In this region, the partitioning of solute atoms at the solid-liquid interface is required as indicated by the dashed arrow in the simplified Al-Si phase diagram (Fig. 2a). Therefore, the interfacial velocity $v_i$ is controlled by solute diffusion at this stage. When the interface further cools down below the solidus, the system enters the single-phase (α-Al phase) region, and the partitioning of solute is no longer required. Consequently, the interface becomes thermal diffusion controlled just like the growth of thermal dendrites of pure metals. Due to the huge difference between the thermal diffusivity and solutal diffusivity (typically about $\times 10^4$ for metals and alloys), the interfacial velocity is expected to be quite different in these two regions.

To further test our hypothesis, we followed the LKT analysis[44] and explored the dependence of interfacial velocity on undercooling $\Delta T$. Fig. 2(b) shows the log-log plot of velocity vs normalized undercooling ($\Delta \bar{T} = \Delta T/\Delta T_0$) for Al-10Si calculated using the LKT model. (The freezing range $\Delta T_0$ here and afterwards is specifically defined as the difference between the liquidus and solidus of the $\alpha$ phase.) The log-log plot shows a linear dependence for small undercooling until the undercooling gets closer to the freezing range $\Delta T_0$, where a clear increase is observed. To further illustrate how the velocity change with the undercooling near the solidus, we plotted $dv/dT$ vs $\Delta T$. As shown in Fig. 2(c), there is a dramatic jump near the solidus. This anomaly of velocity change near the solidus applies to metals in general, as analyzed in the Supplementary Materials. Thus, the above analysis confirms our hypothesis that the sudden velocity jump corresponds to the crossing of the solidus, and it is due to the switching of the controlling mechanism from solutal diffusion-controlled to thermal diffusion-controlled.

It should be noted that both the simplified Al-10Si phase diagram (Fig. 2a) and the phase-field model extrapolated the solidus linearly. This is a typical treatment for modeling and a reasonable approximation for small extrapolations. However, as shown in Fig. 2a, the extrapolated solidus of Al-10Si is about 400 K, which is about 450 K below the equilibrium eutectic temperature and should not be the truth. Therefore, our predictions on the transition temperature for controlling mechanism switching should not be considered as accurate. With more thermodynamic data on the extrapolated solidus, the prediction can be improved. But our conclusions on the controlling mechanism switching and the microstructure development still stand.

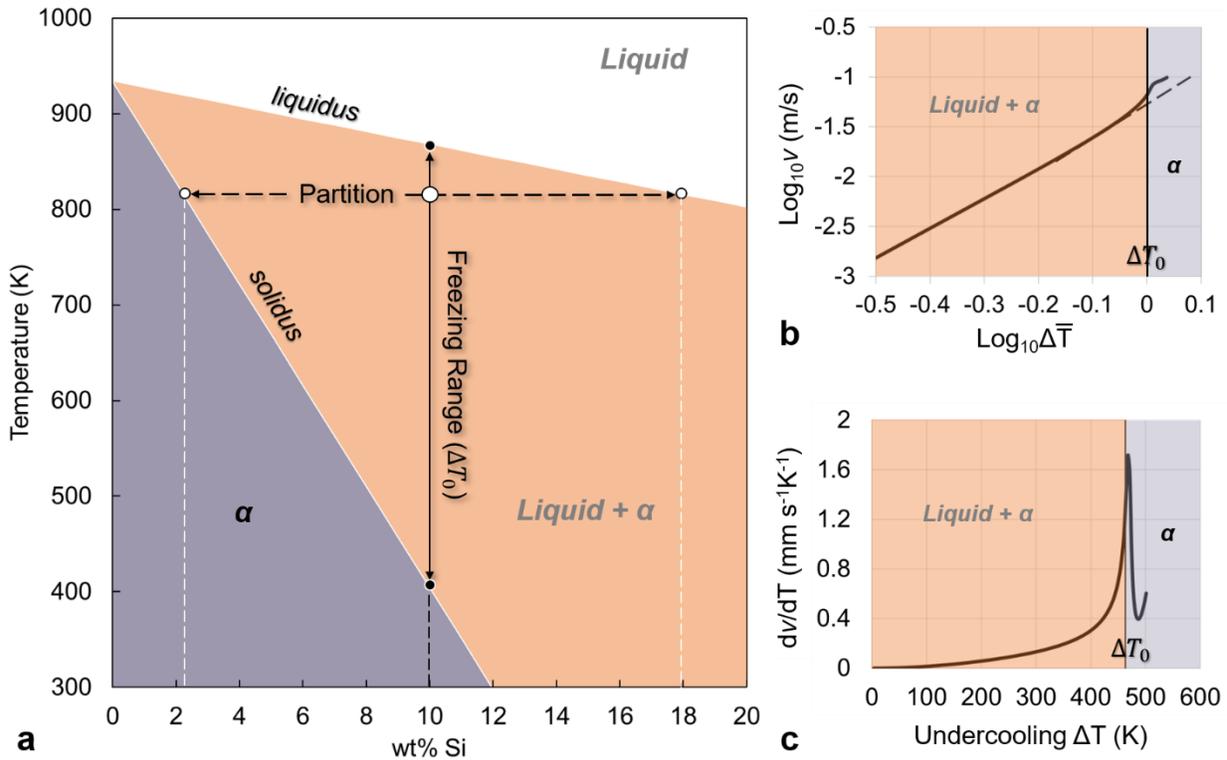

Figure 2. The switching of controlling mechanisms during the rapid solidification process using Al-10Si alloy as an example. a. Al-Si phase diagram at Al rich corner. The solidus is extended to 300 K to illustrate the freezing range ($\Delta T_0$). b. The log-log plot of interfacial velocity vs normalized undercooling ($\Delta \bar{T} = \Delta T/\Delta T_0$) for Al-10Si. In the two

phase-region $\log v$ depends on $\log \Delta \bar{T}_0$ almost linearly. (c) The relationship between $dv/dT$ and undercooling $\Delta T$ for Al-10Si. Near the solidus temperature ($\Delta T = \Delta T_0$), $dv/dT$ reaches its maximum.

## 2.3 The Initial Transient and the Steady State

A typical solidification process can be divided into initial transient, steady state, and the final transient, assuming limited liquid diffusion and no convection[43]. The solidification begins with initial solid forming of composition $kc_\infty$, where $k$ is the partition coefficient. As the solidification continues, a steady state is approached, and the solid composition is exactly the overall composition $c_\infty$. The simulated solidification process of the melt pool of Al-10Si shows similar characteristics in its first two stages (see the top panel of Fig. 1f). Because our model does not consider secondary solid phases, the final transient cannot be properly reproduced. Therefore, we focus on the first two stages.

In this subsection, we use Al-4Cu as an example to investigate the initial transient and steady state stage. The major reason we chose Al-4Cu other than Al-10Si is that no extrapolation of solidus is required. In addition, most of the parameters of rapid solidification for Al-Cu are measured experimentally. In the following, we perform phase-field simulations to demonstrate that 1) the initial transient and steady state are typical for the rapid solidification of a melt pool during AM; 2) the solutal diffusion-controlled first stage corresponds to the initial transient; and 3) the thermal diffusion-controlled second stage corresponds to the steady state.

During the AM metal processing, it is generally believed that the interfacial velocity or the growth rate is dictated by the moving speed of the energy source, which indicates the solidification is thermally controlled. According to our analysis above, the system should already be in the single-phase region. However, when the solidification begins, the interfacial velocity is zero and the temperature is just below the liquidus. Therefore, there must exist an initial transient stage for the velocity to increase to catch the moving speed of the energy source and for the temperature to drop low enough such that the motion of the interface can be thermally controlled. This stage is solutal diffusion controlled as we discussed in the previous subsection and is evident in our phase-field simulations (Fig. 1f and simulations below).

In Fig. 3, we show the phase-field simulation results of the rapid solidification of melt pools of Al-4Cu with different cooling rate applied to the melt boundary. Instead of interfacial thermal conduction boundary condition, we fixed the cooling rate to $10^6$, $5 \times 10^6$, $2 \times 10^7$, and $5 \times 10^7$ K/s at the boundary of the melt pool to test if the interfacial velocity is thermal diffusion controlled. Fig. 3(a) shows a clear correlation between the cooling rate and the velocity. The higher the cooling rate is, the lower the interfacial temperature becomes, and consequently the higher interfacial velocity is in the second stage. This clearly indicates a thermally controlled mechanism for the second stage. In addition, the temperature profiles show recalescence for low cooling rate at the end of the initial transient similar to Al-10Si. At higher cooling rate, $> 2 \times 10^7$ K/s, no clear recalescence is observed.

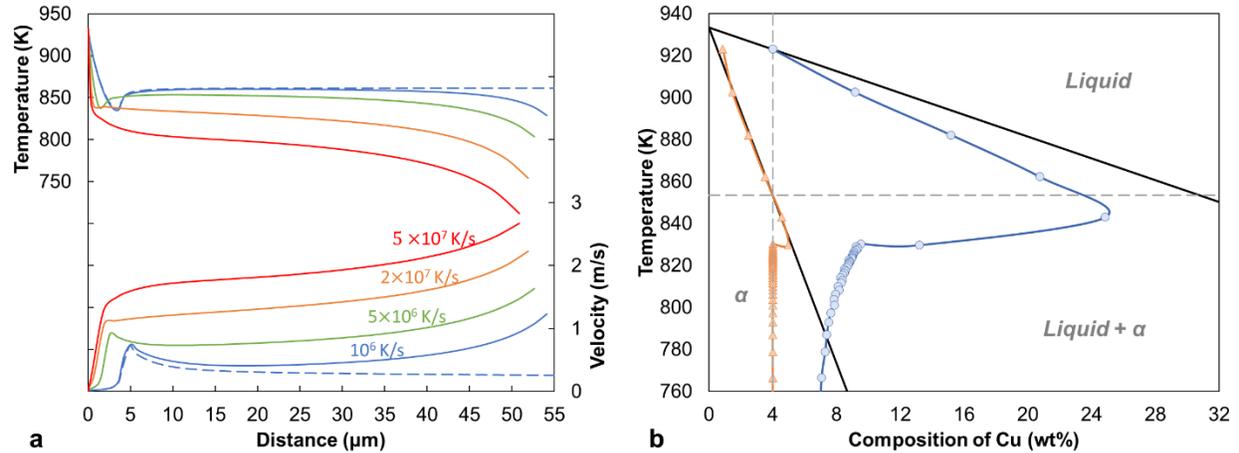

Figure 3. The evolution of temperature, velocity, and solute concentration at the solidification front during the rapid solidification of a melt pool in Al-4Cu simulated by the phase-field model. a. Temperature and velocity vs distance from the melt pool boundary. The blue dashed lines are calculated for the cooling rate of $10^6$ K/s using the same simulation setup except for Cartesian coordinates, which accounts for constant interface. b. Solute concentrations on both sides of the solid-liquid interface as the melt pool solidifies with a cooling rate of $2 \times 10^7$ K/s enforced at the melt pool boundary.

Fig. 3a also shows that all the simulated cases have clear two stages. In contrast to the simulation results of Al-10Si, the second stage shows nearly constant velocities in all simulated cases, which is due to fixed cooling rates at the melt pool boundary. The slight increase in the velocity in this stage, which is also observed in laser remelting Al-Cu experiments[23,45], is due to the decreasing interfacial area of the solidifying hemisphere. If the interfacial area is constant, the temperature and velocity are almost flat in the second stage (blue dashed line in Fig. 3a), which are features of a steady state. Therefore, it can be concluded that the second stage of the rapid solidification of a melt pool is controlled by thermal diffusion.

Fig. 3b shows the evolution of compositions on both sides of the interface for the cooling rate of $2 \times 10^7$ K/s enforced at the melt pool boundary. The solute concentration on the solid side follows the solidus, while the concentration on the liquid side deviates from the liquidus due to solute trapping as the interfacial velocity increases. When the interfacial temperature cools below the equilibrium solidus, the solute concentration becomes higher than the alloy composition and keeps increasing. Eventually, as the interfacial velocity is high enough, the composition of the interface jumps to the alloy composition and the solidification enters the second stage. Then, the solute concentration sticks to the alloy composition and remains unchanged, because no solute partitioning is required.

The highest velocity simulated in this study is about 3 m/s, which is still much smaller than the diffusive velocity $v_D$ (~10 m/s, see Supplementary Materials). The complete solute trapping is therefore not possible. Hence the solute concentration profile shown in Fig. 3b are not following $T_0$ line, which represents the phase boundary between solid and liquid when their respective Gibbs free energies are equal.

**2.4 The Undercooling**

Since the boundary between the initial transient and the steady state is essentially where the interfacial temperature approaches solidus, the undercooling $\Delta T$ at this point is equal to the freezing range $\Delta T_0$. The initial transient can thus be regarded as the stage for the interface to cool down from the liquidus to solidus. Correspondingly, the width of the initial transient is related to the freezing range of the alloy. For a typical AM alloy processing, the cooling rate $\dot{T}$ is ~$10^6$ K/s and the interfacial velocity in this stage (solutal diffusion controlled) is 0.01~0.1 m/s (Fig. 1f and 2b). For a $\Delta T_0$ of 100 K, the width of coarse band (initial transient) is roughly 1~10 μm ($= \frac{\Delta T_0}{\dot{T}} v_i$). As experimentally observed coarse band of AlSi10Mg is around 5~20 μm in width, the freezing range is on the magnitude of ~200 K. Therefore, similar coarse band structures are expected in alloys with large freezing range. For alloy systems with small freezing range, the initial transient stage may be too small to develop any observable microstructure features.

To facilitate the discussion, we define a freezing range parameter (FRP) as the ratio of the freezing range and alloy composition assuming both the solidus and the liquidus are linear dependent on alloy composition. Large FRP indicates the system may develop large coarse band similar to AlSi10Mg with relatively low solute concentration. Table 1 lists the calculated FRP and some other thermodynamic parameters of alloying element in Al[46,47]. Because most of the wrought Al alloys reside at the Al-rich corner of the phase diagram (low solute concentration), they do not form a clear large coarse band during AM. The eutectic Al-Ce alloys were shown to develop complex structures of bands[17] by AM, which will be discussed in details blow. Due to the extremely low solubility of Ce in α-Al, the solidus slope and partition coefficient cannot be determined quantitatively. Hence, the parameter of Ce is not listed here, although this system should have very large FRP. Since the thermal stresses due to thermal expansion/contraction are major contributors to the residual stresses in AM fabricated components, lower solidus should lead to less residual thermal stress. The good AM-printability of eutectic alloys, which usually have high solute concentration, may be partially due to this reason.

For those peritectic elements, which may form Al$_3$X type grain refining primary particles, the FRP indicates how much undercooling the α-Al phase may get when it begins to solidify. As will be discussed in the Al-Zr example below, the smaller the FRP is, the larger the primary Al$_3$X particles need to be for enough nucleation potency.

Table 1. Thermodynamic parameters of alloying element in Al[46,47]

| Element | Petition coefficient $k$ | Liquidus slope $m$ (k/wt%) | freezing range parameter (K/wt%) |
|---|---|---|---|
| Cr | 2.29 | 4.64 | 2.6 |
| Ti | 6.95 | 24.46 | 20.9 |
| Ta | 2.5 | 70 | 42.0 |
| V | 4 | 10 | 7.5 |
| Hf | 2.4 | 8 | 4.7 |
| Mo | 2.5 | 5 | 3.0 |
| Zr | 2.17 | 2.41 | 1.3 |
| Nb | 1.5 | 13.3 | 4.4 |
| Si | 0.124 | -6.59 | 46.3 |
| Ni | 0.043 | -2.51 | 56.6 |
| Mg | 0.375 | -5.24 | 8.7 |
| Fe | 0.025 | -3.48 | 133.5 |
| Cu | 0.133 | -2.61 | 17.0 |
| Mn | 0.724 | -1.16 | 0.4 |
| Sc | 0.665 | -1.35 | 0.7 |

For the AM of hyperperitectic and hypereutectic alloys, the microstructure development is complicated by the primary solid phases and their structures. For instance, the Al-12Ce is a slightly hypereutectic alloy (Fig. 4a). The first microstructure formed during rapid solidification is a pure eutectic structure[17] located at the melt pool boundary, which corresponds to the temperature range between $T_1$ and $T_2$. (The relatively fast velocity suppressed the formation of single $Al_{11}Ce_3$ and favors the formation of eutectic structure. That's why it does not follow the equilibrium phase diagram.[17]) As the interfacial temperature gets lower than the extrapolated liquidus of α-Al ($T_2$), cellular or dendritic α-Al begins to form. Since this is still controlled by solutal diffusion, the formation of α-Al and eutectic structures are both slow, which leads to the observed mixed region next to the eutectic region. As the interfacial temperature gets lower than the extended solidus of α-Al ($T_3$), the interfacial velocity is thermally controlled and becomes much faster than the formation of eutectic structures. Therefore, the microstructure evolves to dendritic/cellular with the eutectic structures decorating the boundaries[17]. The microstructure of a laser-remelted Al-12Ce melt pool is shown in Fig. 4b, which consists of three distinct regions and agrees well with our analysis. The original paper[17,48] applied the microstructure selection model to explain the observed microstructure (Fig. 4b), which is more quantitative than our theory. However, the controlling mechanism switching theory enables us to quickly understand why such microstructure formed and maybe semi-quantitatively relate some microstructure features (such as the width of the band) to velocity and undercooling without performing complex calculations.

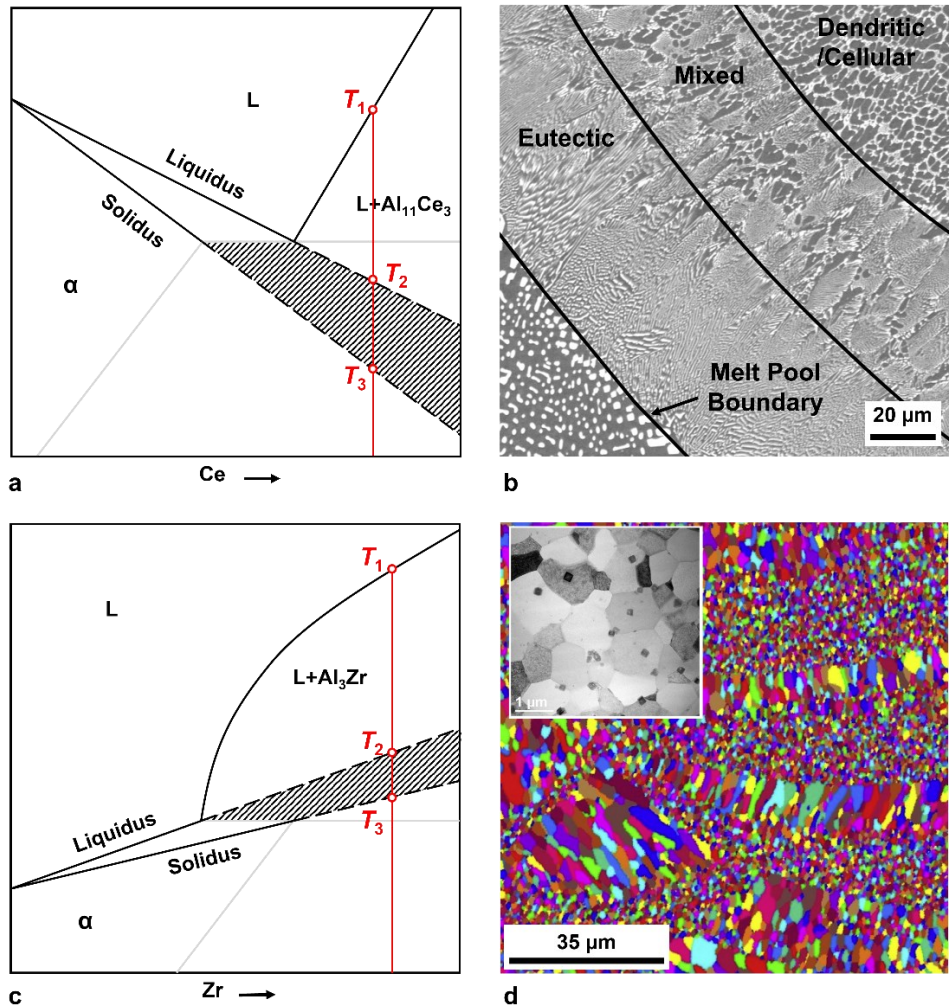

Figure 4. The characteristic microstructures of hypereutectic (Al-12Ce) and hyperperitectic (Al-Zr) alloys developed during the rapid solidification of AM. a. Schematic eutectic phase diagram for Al-Ce (the shaded area is the extended L+α two-phase region); b. microstructure of a melt pool in a laser remelting experiment of Al-12Ce[17]; c. schematic peritectic phase diagram for Al-Zr; d. EBSD map showing the cross section of peak-aged Al-3.60Mg-1.18Zr wt.% alloy fabricated by AM[9]. The inset of d is the low magnification BF-STEM image taken from the fine-grain region showing primary $Al_3Zr$ particles (dark colored cubes) both within the grains and at the grain boundaries[9].

The microstructure developed in hyperperitectic alloys can also be properly explained by the switching of the controlling mechanisms. We chose to investigate the microstructure development in hyperperitectic Al-1.2Zr alloy due to its strong cracking resistance and potential for age hardening[9]. During the first stage of the solidification between $T_1$ and $T_2$ (Fig. 4c), primary $Al_3Zr$ particles nucleate and grow. As the temperature cools below $T_2$, the extrapolated liquidus of α-Al, solid α-Al phase begins to form. Because the $Al_3Zr$ particles are perfect heterogeneous nucleation sites for α-Al, the solidification of α-Al starts on those primary particles. In this stage, α-Al slowly solidifies as this is solutal diffusion-controlled, and new $Al_3Zr$ phase continues to form. When the temperature is lower than $T_3$, the interfacial velocity of α-Al becomes thermally controlled and hence much faster. The region near the melt pool boundary is quickly filled with equiaxed grains, as α-Al already nucleated on primary $Al_3Zr$ particles in this region. It should be noted that not all the primary Al$_3$Zr particles are potent for heterogeneous nucleation. In the beginning of the third stage ($T \leq T_3$), the temperature is just below the extended solidus, which gives an undercooling of 1.56K ($= T_2 - T_3$). The small undercooling can also be understood using the FRP in Table 1, which is 1.3 K/wt% for Zr. For such a small undercooling, the minimum size of nucleating Al$_3$Zr particles is estimated as ~100 nm using the free growth model[49] (see Supplementary Materials). Therefore, in the region far from the melt pool boundary where there are no large enough potent Al$_3$Zr particles, columnar Al grains form. The melt pool eventually consists of fine equiaxed grains near the melt pool boundary and columnar grains filling the rest of the melt pool.

Those analyses were confirmed by AM experiments on Al-Mg-Zr alloys[50]. Experimentally, it was observed that the grain structures of those as-fabricated Al-Zr alloys are bimodal, containing fine equiaxed and coarse columnar grains (Fig. 4d). The equiaxed grains are located next to the melt pool boundary and they nucleated from Al$_3$Zr particles ranging from 100 to 400 nm[50] (inset of Fig. 4d). No nucleating Al$_3$Zr particles less than 100 nm were found, which confirms our prediction on the potent particle size. The coarse columnar grains have no observable Al$_3$Zr particles. The disappearance of Al$_3$Zr particles in columnar region cannot be explained with one single mechanism such as solute trapping. Complete solute trapping is unlikely for typical AM experiments, because the interfacial velocity is not high enough. (See Supplementary Materials) The extended solubility may be the major mechanism. Those smaller Al$_3$Zr primary particles that are not potent for nucleation may dissolve back into the liquid or form Al via peritectic reactions after they were pushed to the grain boundary area. Further investigation is needed to clarify.

The above two examples demonstrated how the switching of the controlling mechanism leads to the formation of heterogenous microstructures across the melt pool. They also indicated that the microstructure development can be manipulated by applying the switching of controlling mechanisms. With the thermodynamic parameters shown in Table 1, alloys with desired primary phases and freezing range to form refined microstructures or layered heterogenous structures can be designed. In addition, the processing conditions can be manipulated to tune the width size of those layers/bands and the size of the dendrites/cells to achieve better mechanical properties such as strength, elongation, fracture toughness, and resistance to cracking.

## 3. Conclusion

In this work, we used a phase-field model incorporated the coupled thermal-solutal diffusion and solute trapping effect to investigate the microstructure development of a whole melt pool during AM. The predicted evolution of the interfacial velocity, solute concentration and temperature explains how the coarse bands formed and why they form in Al-Si alloys. More importantly, the simulation revealed two distinct stages of the rapid solidification for fusion-based metal AM processes, which are controlled by different mechanisms. When the solidification starts, the system lies in the liquid-solid two-phase region of the phase diagram, and the solidification front is controlled by solute diffusion. As the interfacial temperature drops below the solidus and the system gets into the single solid phase region, the solidification front becomes thermal diffusion controlled. This also indicates that the undercooling is larger than the freezing range when the solidification is thermally controlled. Furthermore, the solutal diffusion-controlled region and the thermal diffusion-controlled region correspond to the initial transient and the steady state, respectively, which are the first two stages of a textbook solidification process. The development of those heterogeneous microstructures observed in AlSi10Mg, Al-11.28Si, Al-12Ce and Al-Zr alloys were well explained structurally and compositionally using this controlling mechanism switching theory.

This work is an excellent example of scientific discovery purely guided by modeling. It successfully explained the development of complex heterogeneous microstructures of the entire melt pool for the first time. More importantly, it revealed the switching of the controlling mechanisms happening in a typical solidification process of a melt pool, opening a new horizon for AM alloy design by creating heterogeneous microstructures. This work is a major stepping-stone towards connecting AM processing to microstructure development, which will eventually enable AM to print the desired properties on top of the desired geometries.

**METHODS**

**Phase-Field Modeling**

The phase-field model of solidification for dilute binary alloys with coupled solute and thermal diffusion in one dimension is given by[19,51,52]

$$\tau \frac{\partial \phi}{\partial t} = W^2 \nabla^2 \phi + \phi - \phi^3 - \lambda \ (\theta + Mc_\infty U)(1 - \phi^2)^2 \tag{1a}$$

$$\frac{1+k-(1-k)\phi}{2} \frac{\partial U}{\partial t} = \vec{\nabla} \cdot \left( D_l \frac{1-\phi}{2} \vec{\nabla} U + \vec{J}_{at} \right) + \frac{1+(1-k)U}{2} \frac{\partial \phi}{\partial t} \tag{1b}$$

$$\frac{\partial \theta}{\partial t} = \alpha \nabla^2 \theta + \frac{1}{2} \frac{\partial \phi}{\partial t} \tag{1c}$$

where $\phi$ is the order parameter denoting liquid ($\phi = -1$) and solid phase ($\phi = 1$), $t$ is time, $\theta = \frac{T - T_M - mc_\infty}{L/c_p}$ is the dimensionless undercooling, $T$ is the temperature field, $T_M$ is the melting temperature of pure solvent, $m$ is the liquidus slope of the dilute alloy phase diagram, $c_\infty$ is the nominal concentration of the alloy, $L$ is the latent heat, $c_p$ is the specific heat at constant pressure, $U = \frac{1}{1-k}\left[\frac{2c/c_\infty}{1+k-(1-k)\phi} - 1\right]$ is the dimensionless concentration, $c$ is the concentration of solute atoms, $k$ is the partition coefficient, $\alpha$ is the thermal diffusivity (assumed the same for both liquid and solid), $D_l$ is the solutal diffusivity in the liquid, $\vec{J}_{at} = a_t W[1 + (1-k)U]\frac{\partial \phi}{\partial t}\frac{\vec{\nabla}\phi}{|\vec{\nabla}\phi|}$ is the antitrapping current[53,54], $\lambda = \frac{15L^2}{16Hc_pT_M}$ is a coupling constant, $H = 3\gamma/\delta$ is the height of the energy barrier for the energy double well, $\gamma$ is the interfacial energy, $\delta$ is the interface thickness, $M = -\frac{m(1-k)}{L/c_p}$ is the scaled magnitude of the liquidus slope, $m$ is the liquidus slope of the dilute alloy phase diagram, $\tau = 1/(HM_\phi)$ is the relaxation time, $M_\phi$ is the phase-field mobility related to the movement of the interface, and $W$ is related to the width of the diffuse solid-liquid interface[55] (in this work we chose $W = \delta/2\sqrt{2}$).

To consider the solute trapping effect, $a_t$ in the antitrapping current term $\vec{J}_{at}$ is set to be dependent on $\phi$, i.e. $a_t = \frac{1-A(1-\phi^2)}{2\sqrt{2}}$, where $A = \frac{D_L}{v_D^{PF}W}$ is trapping parameter[33]. $v_D^{PF}$ can be numerically calculated by solving the transcendental relationship between interfacial velocity $v$ and non-equilibrium partition coefficient $k(v)$, [33]

$$k(v) = k_e \exp\left(\frac{\sqrt{2}(1-k(v))v}{v_D^{PF}}\right) \qquad (2)$$

The parameters used in this study is listed in Table 2. The detailed numerical testing, validation, and exploration of interface kinetics, solute trapping effect and thermal diffusion of this model can be found in literature[19].

Table 2. Material properties for Al-10Si and Al-4Cu [46,56]

| Property | Value |
|---|---|
| Melting temperature for pure solvent $T_M$ (K) | 933.47 |
| Equilibrium partition coefficient $k_e$ for Al-10Si | 0.124 |
| Equilibrium partition coefficient $k_e$ for Al-4Cu | 0.13 |
| Equilibrium liquidus slope $m$ (K/wt%) for Al-10Si | -6.59 |
| Equilibrium liquidus slope $m$ (K/wt%) for Al-4Cu | -2.6 |
| Nominal composition $c_\infty$ for Al-10Si (wt%) | 10 |
| Nominal composition $c_\infty$ for Al-4Cu (wt%) | 4 |
| Latent heat $L$ (J/m$^3$) | $9.49 \times 10^8$ |
| Heat capacity $c_p$ (J/k/m$^3$) | $2.81 \times 10^6$ |
| Unit undercooling $L/c_p$ (K) | 336.44 |
| Interfacial energy $\gamma$ (J/m$^2$) | 0.158 |
| Capillary length $d_0 = \gamma T_M c_p / L^2$ (m) | $4.63 \times 10^{-10}$ |
| Solute diffusivity in liquid $D_l$ (m$^2$/s) | $3 \times 10^{-9}$ |
| Thermal diffusivity $\alpha$ (m$^2$/s) | $3 \times 10^{-5}$ |
| Diffusion velocity for solute trapping $v_D^{PF}$ for Al-10Si (m/s) | 1.51 |
| Diffusion velocity for solute trapping $v_D^{PF}$ for Al-4Cu (m/s) | 1.78 |
| Phase-field mobility $M_\phi$ for Al-10Si (m$^3$/J/s) | 1 |
| Phase-field mobility $M_\phi$ for Al-4Cu (m$^3$/J/s) | 10 |
| Interface thickness $\delta$ (m) | $5 \times 10^{-9}$ |

In this study, we focused on the understanding of the fundamental physics of the moving solid-liquid interface and how it leads to the formation of various heterogeneous microstructures during the entire solidification process of a melt pool. Therefore, the overall simulation size should be large enough to cover the whole melt pool and also to account for the super-fast thermal diffusion. Meanwhile, the grid size should be small enough to capture the dynamics of the interface, which include the partitioning of solute, the dissipation of latent heat, and the variation of interfacial velocity. These two requirements lead to a simulation size of at least ~40000 grid points in 1D (for a melt pool of 50 μm) with the grid size of ~1.6 nm. The simulation size is simply squared for 2D, which is extremely challenging in terms of computation cost. We chose to perform 1D simulation of the whole melt pool to capture more physics while sacrificing the curvature effect. Hence, our phase-field model cannot provide direct 2D or 3D morphology information for dendritic or cellular structures. As analyzed in the Supplementary Materials, the curvature effect has limited impact on our simulation results and does not affect our conclusions on the controlling mechanism switching. Therefore, although this modeling work is essentially for planar interface, it applies for other the solidification structures (dendritic and cellular) at least semi-quantitatively. As shown in the Results and Discussion Section, this 1D model successfully captured the essential dynamics of the interface by predicting the velocity evolution, composition variation and thermal history of the entire melt pool, which has never been reported before. Those predictions agree well with experiments (of non-planar structures) and perfectly explained the development of observed microstructures for the first time, which leads to the discovery of the switching of controlling mechanism.

### DATA AVAILABILITY

The data that support the findings of this study are available from the corresponding author upon reasonable request.

### ACKNOWLEDGEMENT

This work was supported in part by the ISC Advanced Manufacturing Signature Area seed fund at Missouri S&T and the National Science Foundation under Grant No. OAC-1919789. Dr. M. Chu and Dr. D. Han are thanked for helpful discussion.

### AUTHOR CONTRIBUTIONS

Y. G. conceived the project, performed the simulations and analysis. Y.G. led the writing of the paper with contributions from all authors. All authors contributed to the discussions.

### COMPETING INTERESTS

The authors declare no competing financial interests.

# Supplementary Materials

## 1. The curvature effect and the rationale of 1D simulations

In this work, we performed 1D phase-field simulations to investigate the rapid solidification process of the whole melt pool during AM. The anisotropy of the interface was not considered as this is a 1D model. The results can be considered as drawn from planar interface assumption. So, the question is how big the effect of the interface anisotropy is and whether our finding is still valid for dendritic or cellular structures when interface anisotropy is considered.

The effect of interfacial anisotropy, or the curvature effect, can be quantified as the curvature undercooling. We followed the widely used LKT analysis[1] to calculate the overall undercooling and its constituents, which include thermal undercooling $\Delta T_t$, constitutional undercooling $\Delta T_c$, and curvature undercooling $\Delta T_r$. The parameters used in this calculation are listed in Table 2 of the main text.

Fig. S1 shows the relative undercooling of each constituent of Al-10Si, which are normalized by the unit undercooling ($L/c_p$). The curvature undercooling $\Delta T_r$ has moderate contribution to the overall undercooling when the velocity of the interface is low or when the overall undercooling is low. Therefore, even though the curvature effect is not considered, our simulations agree quite well with experimental observations, which are non-planar structures, as demonstrated in Section 2.1 of the main text.

As the overall undercooling approaches the freezing range $\Delta T_0$ and beyond, its contribution diminishes and becomes negligible. This trend is generic for metals as illustrate in Fig. 2(b) of the original LKT analysis paper[1]. Hence, the curvature effect does not affect our findings that the controlling mechanism switches at the solidus temperature.

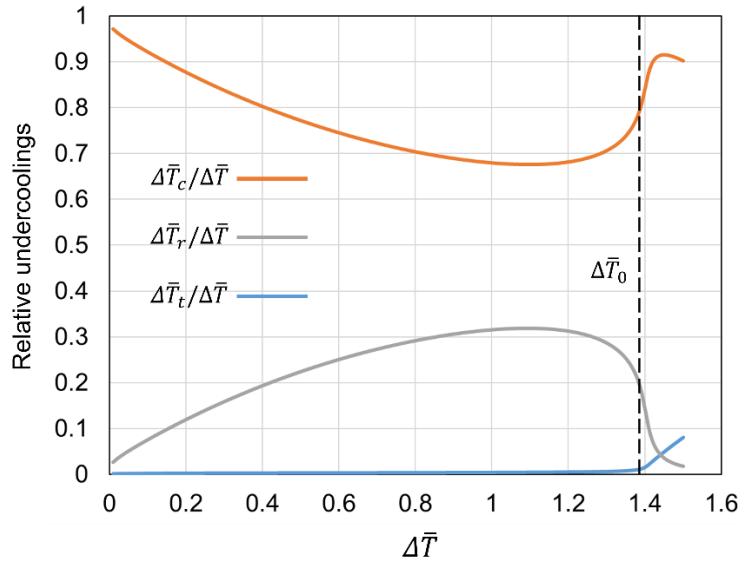

Figure S1. Relative undercoolings as a function of dimensionless undercooling ($\Delta \bar{T}$) for Al-10Si. $\Delta \bar{T}_0$ is the freezing range.

The solidification of the whole melt pool is rather complicated. The thermal diffusion length during the solidification process of a melt pool is on the magnitude of $\sim 10^{-4} m$, assuming the solidification time is $\sim 1\ ms$, and the thermal diffusivity is $10^{-5}\ m^2/s$. Therefore, to correctly describe the evolution of the temperature profile, the simulation needs to cover the whole melt pool (at least ~50 μm) for the whole solidification process. On the other hand, to capture the interplay between solute concentration, temperature, and velocity, the maximum simulation grid size is required to be smaller than the interfacial thickness (~5 nm). Therefore, a ~40000 × 40000 simulation size is required for 2D simulation, which is challenging for even well paralleled computer codes. As demonstrated above, the contribution of interfacial anisotropy is limited to low velocity regime and will not change our major conclusions. Hence, we chose 1D simulation and neglected the curvature effect to capture the essence of the solidification of the whole melt pool. Although our simulation results should be regarded as predictions under planar interface assumption, they are sufficient to reproduce the major experimental observations and reveal how those observed microstructures were developed at least semi-quantitatively.

## 2. Mathematical analysis on the LKT model

In this section, we show that mathematically the velocity should always have a jump near the solidus or when undercooling approaches freezing range based on the analytical LKT model.

According to Lipton *et al*[1], the normalized total undercooling

$$\Delta \bar{T} = \mathrm{Iv}(P_t) + \bar{C}_0(A-1) + 2\sigma^* P_t F(A) \tag{S.1}$$

where $\mathrm{Iv}(P) = P \exp P E_\mathrm{I}(P)$ is the Ivantsov function, $F(A) = \xi_t + \xi_c 2\eta \bar{C}_0 A(1-k_0)$, $A = \frac{1}{1-(1-k_0)\mathrm{Iv}(P_c)}$. $P_c$ and $P_t$ are solutal and thermal Péclet number, respectively. They are related by $P_c = \eta P_t$, where $\eta$ is the ratio of thermal/solutal diffusivity.

As shown in Fig. S2, near the freezing range ($\Delta \bar{T}_0$), $P_t$ jumps from 10$^{-3}$ to 10$^{-2}$. Because $\eta$ is typically on the magnitude of 10$^4$ for metals, $P_c$ goes from 10 to 10$^2$. Therefore, when the undercooling approaches freezing range, $P_c$ is large, but $P_t$ is still small. Hence, it can be approximated that $\xi_c \cong \frac{1}{4\sigma^* k_0 P_c^2}$ and $\xi_t \cong 1 - P_t\sqrt{\sigma^*}$. The total undercooling becomes

$$\Delta \bar{T} = \mathrm{Iv}(P_t) + \bar{C}_0(A-1) + 2\sigma^* P_t\left(1 - P_t\sqrt{\sigma^*}\right) + \frac{\bar{C}_0 A(1-k_0)}{k_0 P_t \eta} \tag{S.2}$$

Because $\mathrm{Iv}(P_c) \cong 1$ when $P_c$ is large, it can be approximated that $A \cong 1/k_0$ and $\bar{C}_0(A-1) \cong \Delta \bar{T}_0$. In addition, the third term on the right-hand side is approximately zero, since $\sigma^* = 1/4\pi^2$ and $P_t$ is small. Then, we have

$$\Delta \bar{T} = \frac{1}{P_t}\left(1 + \frac{\Delta \bar{T}_0}{k_0 \eta}\right) + \mathrm{Iv}(P_t) + \Delta \bar{T}_0 \tag{S.3}$$

Taking the derivative, we arrive at

$$\frac{dP_t}{d\Delta \bar{T}} = \frac{P_t}{P_t(\mathrm{Iv}(P_t)-1)-(\Delta \bar{T}-\Delta \bar{T}_0)} \tag{S.4}$$

Thus, $\frac{dP_t}{d\Delta\bar{T}}$ becomes large when $\Delta\bar{T}$ approaches $\Delta\bar{T}_0 + P_t(\text{Iv}(P_t) - 1)$, which is roughly $\Delta\bar{T}_0$. It can be inferred that $P_t$ should have a jump when undercooling $\Delta\bar{T}$ approaches the freezing range $\Delta\bar{T}_0$. Since the normalized velocity $\bar{V}$ is related to $P_t$ via the following relationship,

$$\bar{V} = \sigma^* P_t^2 F(A) = \frac{1}{2}\left[1 + \frac{\bar{C}_0 A(1-k_0)}{k_0 \eta}\right] = \frac{1}{2}\left(1 + \frac{\Delta\bar{T}_0}{\eta}\frac{1}{1-(1-k_0)\text{Iv}(\eta P_t)}\right) \tag{S.5}$$

$\bar{V}$ is expected to have a jump near $\Delta\bar{T}_0$ as well.

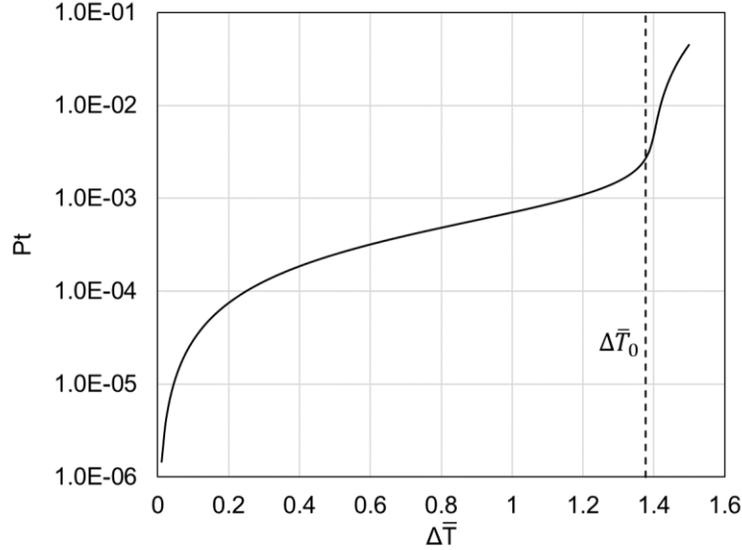

Figure S2. $P_t$ as a function of normalized undercooling $\Delta\bar{T}$ for Al-10Si. $\Delta\bar{T}_0$ is the freezing range.

### 3. The critical size of primary Al$_3$Zr particles

Based on the free growth model[2] and assuming the undercooling is low (1.0 K), the critical size of primary Al$_3$Zr particles is

$$d = \frac{\gamma}{\Delta S \Delta T} = \frac{0.158 \times 933.47}{9.468 \times 10^8 \times 1} = 155.8 \text{ (nm)}$$

### 4. Analysis on complete solute trapping

For rapid solidification, the assumption of local equilibrium breaks down, and solute trapping happens[3]. The partition coefficient $k$ becomes a function of velocity[4,5]. For dilute alloys, the dependence is given by[6]

$$\begin{cases} k = \dfrac{k_e\left[1-\left(\frac{v_i}{v_D}\right)^2\right] + \frac{v_i}{v_D}}{1-\left(\frac{v_i}{v_D}\right)^2 + \frac{v_i}{v_D}}, & v_i < v_D \\ k = 1, & v_n \geq v_D \end{cases} \tag{S.6a}$$

where $k_e$ is the equilibrium partition coefficient, $v_D = D_l/a$ is the diffusive velocity, and $a$ is the atomic jump distance in the liquid. For small velocity ($v_n \ll v_D$), the partition coefficient can be approximated as [5]

$$k = (k_e + \frac{v_i}{v_D})/(1 + \frac{v_i}{v_D}) \tag{S.6b}$$

For metals and alloys, the atomic jump distance ($a$) is on the magnitude of $10^{-10}$ m and the solutal diffusivity is on the magnitude of $10^{-9}$ m²/s. Therefore, the diffusive velocity $v_D$ is on the magnitude of 10 m/s. The typical velocity of AM energy source is on the magnitude of 0.1 - 1 m/s. Therefore, complete solute trapping is rare for typical AM metal processes, and the approximation (Eq. S.6b) usually applies.

5. **Difficulty in experimental validation**

Due to the fast-moving interface and small interface thickness, it is very challenging to obtain experimental validation for the interface temperature. Below we show an analysis on capturing the temperature increase (recoalescence) experimentally to validate our phase-field predications as an example.

Figure S3 shows the phase-field simulation of temperature evolution at different locations in a melt pool of Al-4Cu during the whole solidification process. Nx is the total simulation length corresponding to the radius of the whole melt pool. The cooling rate imposed at the melt pool boundary is $10^6$ K/s, corresponding to the light blue curve in Fig. 3a of the main text. The recalescence (temperature increase) takes ~40 µs in every location, and the increased temperature is about 20 K.

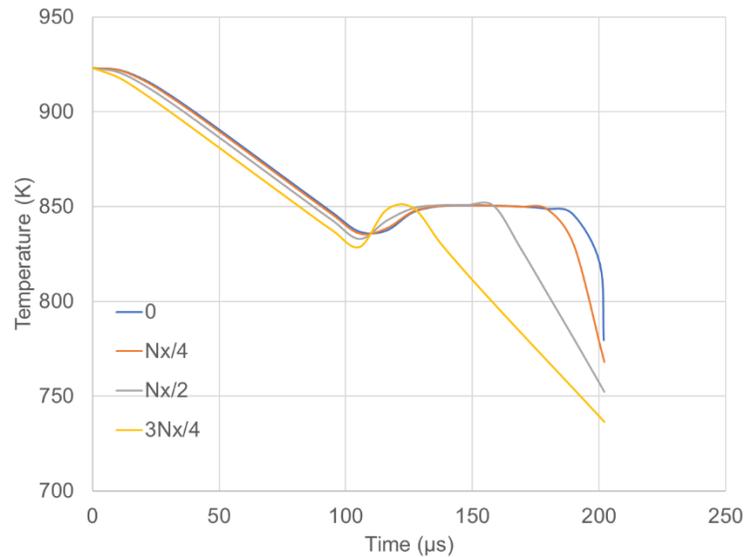

Figure S3. The temperature evolution at different locations of the melt pool for Al-4Cu with fixed cooling rate of $10^6$ K/s imposed at the melt pool boundary. (The simulation corresponds the light blue curve in Figure 3 of the main text.) Nx is the total simulation length. 0 corresponds to the center of the melt pool. Nx/4, Nx/2 and 3Nx/4 correspond to quarter, half and three quarters of the melt pool length away from the center of the melt pool.

There are two experiments we can perform to possibly capture this recalescence.

1. *Using a thermal camera to capture the temperature directly*. The thermal camera we can access has a spatial resolution of 10 μm and frame rate of 100 kHz, which is among the most advanced. During the whole recalescence, the camera can take ~ 4 images. The temperature showed in this figure is calculated of a point of about 1.4 nm (smaller than the interface). Due to the resolution limit, the captured temperature by the thermal camera should be averaged over 10 μm, which leads to a much smaller temperature difference than 20 K shown in Fig. S3. Therefore, both spatial and temporal resolution are not adequate to capture this phenomenon.

2. *Using in-situ X-ray imaging to capture the velocity decrease to indirectly infer the temperature increase*. The state-of-art in situ X-ray imaging has a spatial resolution of 2 μm and a frame rate of 1 MHz. To capture the solidification front velocity, a decent travel distance, e.g., 10 μm, is needed. However, the distance the interface moved with a clear decreasing velocity during the recalescence is about 2 μm as shown Fig. 3a (blue curve for velocity evolution) of the main text. It is thus not sufficient to detect the velocity variation.

In summary, neither of these experimental techniques is good enough to capture the recalescence. In fact, from our experience of observed velocity change using in situ X-ray imaging, no velocity decrease was ever observed, which is consistent with the above analysis. In addition, because of the limitations of experiments, the theoretical work becomes very important since it can provide exclusive insights into those phenomena.